\documentclass[12pt,preprint]{aastex}
\input{epsf}
\received{}
\accepted{}

\shorttitle{Exohost2}
\shortauthors{Mason et al.}

\begin{document}
\slugcomment{to be submitted to {\it The Astronomical Journal}, \\
Version: 9/08/2011}
\title{Know the Star, Know the Planet. II. \\
Speckle Interferometry of Exoplanet Host Stars}

\author{Brian D. Mason\altaffilmark{1}, William I. Hartkopf\altaffilmark{1}}
\affil{United States Naval Observatory\\3450 Massachusetts Ave., NW, 
Washington, DC 20392-5420\\Electronic mail: (bdm,wih)@usno.navy.mil}

\author{Deepak Raghavan\altaffilmark{1}}
\affil{Georgia State University, Dept. of Physics \& Astronomy, Atlanta, GA 
30303-3083\\Electronic mail: raghavan@chara.gsu.edu}

\author{John P. Subasavage\altaffilmark{1}}
\affil{Cerro Tololo Inter-American Observatory, La Serena, Chile
\\Electronic mail: jsubasavage@ctio.noao.edu}

\author{Lewis C. Roberts, Jr.}
\affil{Jet Propulsion Laboratory, California Institute of Technology, 4800 
Oak Grove Drive, Pasadena CA 91109\\Electronic mail: 
lewis.c.roberts@jpl.nasa.gov}

\author{Nils H. Turner and Theo A. ten Brummelaar}
\affil{Center for High Angular Resolution Astronomy, Georgia State 
University, Mt.\ Wilson, CA 91023\\Electronic mail: 
(nils,theo)@chara-array.org}


\altaffiltext{1}{Visiting Astronomer, Kitt Peak National Observatory
and Cerro Tololo Inter-American Observatory, National Optical
Astronomy Observatories, operated by Association of Universities for
Research in Astronomy, Inc.\ under contract to the National Science
Foundation.}

\begin{abstract}



A study of the host stars to exoplanets is important to understanding their 
environment. To that end, we report new speckle observations of a sample of 
exoplanet host primaries. The bright exoplanet host HD 8673 (= HIP 6702) 
is revealed to have a companion, although at this time we cannot 
definitively establish the companion as physical or optical. The observing
lists for planet searches and for these observations have for the most part 
been pre-screened for known duplicity, so the detected binary fraction is 
lower than what would otherwise be expected. Therefore, a large number of 
double stars were observed contemporaneously for verification and quality 
control purposes, to ensure the lack of detection of companions for 
exoplanet hosts was valid. In these additional observations, ten pairs are 
resolved for the first time and sixty pairs are confirmed. These observations 
were obtained with the USNO speckle camera on the NOAO 4m telescopes at both 
KPNO and CTIO from 2001 to 2010.

\end{abstract}


\section{Introduction}

As discussed in Paper 1 of this series (Roberts et al.\ 2011) a study of the 
host stars to exoplanets is essential if we wish to understand the 
environment in which those planets formed. Further, the star's luminosity,
distance, mass, and other characteristics are fundamentally related to the
determination of the planets mass and size. Determining these parameters
directly for the host star as opposed to using a template of the canonical 
stellar class and type will produce more accurate and precise planetary
determinations. As part of this effort we herein report new speckle 
observations of a large sample of exoplanet host primaries. 

Binaries affect the formation and stability of planetary systems, as their
long-term relationship must be hierarchical. Generally speaking, based on 
the precepts of Harrington (1981) if the ratio of semimajor axes is 4:1 or
greater, an exoplanet in a stellar binary is dynamically stable. Dynamically
permitted systems include the more commonly detected configuration of 
planet(s) orbiting one stellar component of a sufficiently wide-orbit binary
in a hierarchical arrangement, and the harder-to-detect circumbinary 
configuration of planet(s) in a wide orbit around a close stellar binary 
(see Raghavan et al.\ 2006, especially \S 6.1). That said, Raghavan et al.\
(2010) in their statistics updating and improving upon Duquennoy \& Mayor 
(1991), find that while the frequency of single stars is the same, the
number of companions has increased through instrumental and technique 
enhancements. Due to the presence of stellar companions, one might imagine
the environment of binary stars to be a rich one for substellar companions.
However, dynamical effects should eject companions not found in hierarchical
orbits. In any event, as conducive as this environment might be to 
companions, this is not reflected in the list of known planet hosts, 
however. This is due entirely to selection effects; because of the 
complexities of disentangling stellar companions from small planetary 
signatures, the observing lists for planet searches have for the most part 
been pre-screened for known duplicity, so the detected binary fraction is 
lower than what would be expected.

In addition to binary stars that are gravitationally bound, there are 
optical doubles which are merely chance alignments of unrelated stars.
Although they do not contribute dynamically to the system, close optical 
pairs do contribute light to the system, which should be accounted for in 
system analysis. While photometric analysis of binary systems can infer 
$``$third" light in the system, radial velocities or periodic variation in
other astrometric parameters (for example, the Hipparcos acceleration 
solutions), would not give evidence of these companions. Optical pairs 
would best be found by direct imaging or interferometric analysis.

\section{Speckle Observations}

All of these observations were obtained as part of other observing projects,
for example, analysis of white, red and subdwarfs (Jao et al.\ 2009), G 
dwarfs (\S 5.3.6 of Raghavan et al.\ 2010), or Massive stars (Mason et al.\ 
2009), some of which are still in developmental and/or data collection 
stages. Unpublished observations of exoplanet host stars were extracted from
these data and are presented here. The instrument used for these speckle 
observations was the USNO speckle interferometer, described most recently in
Mason et al.\ (2009). 

Speckle Interferometry is a single filled-aperture interferometric technique
where the $``$speckles" of a pair of nearby stars, induced by atmospheric 
turbulence, constructively interfere. Reduced by simple autocorrelation 
methods, in the resulting image the binary or double star geometry is the 
predominant structure when compared with the other uncorrelated pairings.
It is capable of resolving pairs to the resolution limit of the telescope in
question up to size of the observation field (typically, $\sim$1\farcs5), as
long as the pairs have magnitude differences of less than about three.

Calibration of the KPNO data was accomplished through the use of a 
double-slit mask placed over the ``stove pipe'' of the 4-m telescope during 
observations of a bright known-single star (as described in Hartkopf et al.\
2000). This application of the Young's double-slit experiment allowed 
determination of scale and position angle zero points without relying on 
binaries themselves to determine calibration parameters. Multiple 
observations through the slit mask (during five separate KPNO runs from 2001
to 2008) yielded mean errors of 0\fdg11 in the position angle zero point and
0.165\% in the scale error. These ``internal errors'' are undoubtedly 
underestimates of the true errors for these observations, because these 
calibration tests were made on stars that were brighter and nearer the
zenith than science stars. Total errors are likely 3--5$\times$ larger than 
these internal errors.

Because the slit-mask option is not available on the CTIO 4-m telescope, we 
calibrated the Southern Hemisphere data using observations of numerous 
well-observed wide equatorial binaries obtained at both the KPNO and CTIO 
telescopes. Published orbital elements for these pairs were updated as 
needed, using the recent KPNO and published measures, then predicted $\rho$ 
and $\theta$ values from those orbits deemed of sufficiently high quality 
were used to determine the CTIO scale and position angle zero points. The 
calibration errors for these southern observations were (not surprisingly) 
considerably higher than those achieved using the slit mask. Mean errors for
five CTIO runs from 2001 to 2010 (as well as a 2001 KPNO run lacking slit 
mask data) were 0\fdg67 in position angle and 1.44\% in scale. These errors 
are likely overestimated, because we have assumed that the calibration 
binaries have perfect orbits, and any offsets are incorporated into the 
errors.

\section{Results}

Following generation of the Directed Vector Autocorrelation (Bagnuolo et 
al.\ 1992), the $``$DVA" is background subtracted through boxcar subtraction
and the sharp central peak which corresponding to the zeroth order speckles 
correlating with themselves is clipped. Companions in the resulting 
DVA are then readily apparent as peaks several sigma above the background.

Of the 118 exoplanet hosts we observed only one, HIP 6702 showed signs of a 
companion and is discussed in Section 3.1 and listed in Table 1. The null 
results are listed in Table 2, a list of single star detections. In the 
table, Column (1) gives the Hipparcos number, Column (2) the HD Catalog 
number, Column (3) lists other common designations, Column (4) is the epoch 
of observation, and Column (5) identifies the telescope (C = Cerro Tololo 
4m, K = Kitt Peak 4m). For all of these observations no companion was 
detected within the ranges 
$\Delta$m$_{\rm V}$~$<$~$3$, and $0\farcs03~<~\rho~<~1\farcs5$. 

Table 1 lists the observations for this new detection. Column (1) gives the 
precise position of the system, Column (2) is the Washington Double Star 
Catalog (hereafter, WDS; Mason et al.\ 2001) identifier, and Column (3) 
lists the discovery designation, here the WSI (= Washington Speckle 
Interferometry) number. Column (4) gives the Hipparcos number of the primary
as a cross-reference. Column (5) gives the epoch of observation, and Columns
(6) and (7) provide the relative astrometry. Column (8) lists a crude 
estimate of the magnitude difference of the pair in the V band (the listed 
number is paired with the more reliable observation). This estimate is 
probably only good to $\pm0.5\,mag$. Column (9) provides the separation in 
astronomical units, based on the Hipparcos parallax and this angular 
separation assuming a face-on orbit. 

The resulting multiplicity fraction is extremely low, but artificially so.
Observation of known binaries is a prime goal of the USNO speckle program
and some of these pairs had been previously published (e.g., HD 28305 in
Mason et al.\ 2009). Others which were known but whose motions were not 
especially compelling (e.g., HD 50583 in Mason et al.\ 2011) were observed 
with our 26in refractor in Washington and those which do have a compelling
individual story to tell unrelated to exoplanets are in preparation 
(Farrington et al.\ 2012). A simplistic multiplicity determination of this 
limited result ($=~\frac{1}{118}$) is therefore not a meaningful number.

\subsection{New Double Star : HIP 6702}

Of all the exoplanet hosts which have been serendipitously observed, all 
were unknown as close visual doubles and only one of the host stars, HIP 
6702 (= HD 8673) appeared double in directed vector autocorrelations on both
times it was observed. The classification of HIP 6702 as an exoplanet is 
based upon Hartmann et al.\ (2010) who, using iodine-cell radial velocity 
measurements, detected a companion with a $M sin~i$ of 14.2 M$_j$ with a 
period of 1634$\pm$17 days and an eccentricity of 0.723$\pm$0.016. The 
relative astrometric measures of this resolved pair are provided in Table 1.
Given the small number of measures presented in Table 2, the pair, while a 
visual double star, is not necessarily a binary system. Verification of 
physicality for the new companion to HIP 6702 can be accomplished several 
ways, among them color-magnitude, proper motion and/or kinematic analysis. 
The speckle interferometry observable of relative position establishing 
kinematic-physical (i.e., Keplerian) motion requires at least three 
measures. So, while close proximity can be a powerful argument for 
physicality, it is by no means definitive (c.f., $\iota$ Ori, \S 5.1 of 
Mason et al.\ 2009). Nevertheless, even a companion which is only nearby in 
the angular sense should be considered in any detailed analysis of the star,
as it will contribute to the photometric signature of the examined target. 
Such is the case for HIP 6702, which was recently reported as a sub-stellar 
companion (Hartmann et al.\ 2010). 

Among the possible interpretations of the new speckle companion two stand 
out: first, the companion is a not-physically associated line-of-sight 
companion and second, it is the companion detected in Hartmann et al.\ 
(2010).

\subsubsection{Physical Companion?}

Hipparcos produced many types of double star solutions. The one which can be
most easily compared to other detection techniques and the most common are 
those where the relative parameters ($\rho$, $\theta$) are presented. The
speckle interferometry measures presented in Table 1 are both near the 
Hipparcos $``$C" code double star solution cutoff (0\farcs082 for HDS 446 = 
HIP 27151). The other Hipparcos double star solutions may not be applicable 
here. Some depend upon {\it a priori} orbital information (O code), system 
dynamics in the plane of the sky (G code), variability (V code) or unknown 
parameters (X and/or S code). In any event, the lack of Hipparcos detection 
is a condition which is neither necessary nor sufficient to establish that 
the Hartmann et al.\ (2010) companion is not-stellar. 

However, if the two Table 1 measures represent relative measures of the 
Hartmann et al.\ (2010) pair, the inclination must be extremely low. 
Assuming a near zero inclination the mean separation of 0\farcs098 would 
approximate the angular semi-major axis (a$''$ = 0\farcs098$\pm$0\farcs011).
Given this, the Hipparcos parallax of 26.14$\pm$0.79 mas and the Hartmann et
al.\ period of 1634$\pm$17 days, a mass sum of 2.63$\pm$0.92 M$_{\odot}$ is
obtained, which is not unreasonable for two similar F dwarf stars, although 
the error is quite large, primarily due to the uncertainty in a$''$. The 
length of time between the two speckle observations represents 1.47$\times$ 
the Hartmann et al.\ period. The two measures of angular position represent 
$(0.497~or~0.503)+n$ revolutions of the system (depending on direction of 
rotation) which is very similar to the Hartmann et al.\ period when $n=1$.

Given the estimated dynamic range ($\Delta$m$_{\rm V}$ = $2.3\pm0.5$) and
assuming the fainter limit and spectral type of the primary this would make 
the secondary close to a K2V. Using the canonical mass of a K2V in 
$M~sin~i=14.2M_j$ gives an inclination of 1\fdg02. Using this inclination 
with $a~sin~i$ from Hartmann et al.\ (2010) gives a semi-major axis of 
0\farcs168 which is consistent with the Table 1 results.

\subsubsection{or Optical Companion?}

Since the interferometric companion to HIP 6702 has been observed so few 
times, establishing the companion as optical or physical is not possible. 
The proper motion of the primary is 0\farcs25/yr ($\alpha$=0\farcs236/yr,
$\delta$=$-$0\farcs085/yr). From the relative positions in Table 1, the 
proper motion of the companion would be an even higher at 0\farcs276/yr. 
If linear motion is assumed and reasonable errors are applied it is possible
to determine where the companion would be at some date in the future. In 
Figure 1 this determination is performed assuming errors slightly larger 
than nominal for the two speckle interferometry measures: $\Delta\theta = 
1\fdg0$, $\Delta\rho/\rho = 1.0\%$ The predicted position for 2012 through 
2015 are plotted as error boxes. Again, assuming linear motion from the two
speckle points, a separation of 0\farcs37 and a position angle of 
255$^{\circ}$ is determined for 1991.25, which would be well within the 
capabilities of the Hipparcos satellite (ESA 1997).

\begin{figure}[h]
\begin{center}
{\epsfxsize 4.0in \epsffile{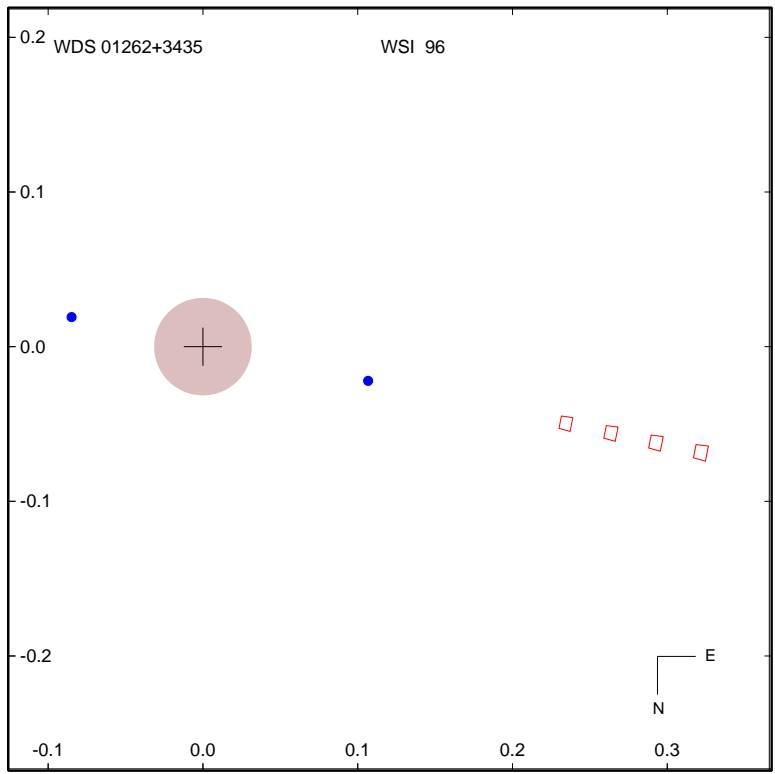}}
\end{center}
\caption{\small A ``motion characterization'' system plot for HIP 6702 (= 
WSI~~96) with small filled circles indicating the 2001 and 2007 
speckle measures from Table 1. Scales are in arcseconds, and in each figure 
the large shaded circle represents the V band resolution limit of a 4m 
telescope. The four small error boxes in each figure indicate the predicted 
location of that pair's secondary in 2012.0, 2013.0, 2014.0, and 2015.0, 
assuming the motion is linear and all speckle measures are characterized by 
errors of $\Delta\theta = 1\fdg0$, $\Delta\rho/\rho = 1.0\%$. Finding the 
double within a box appropriate to the observation date would be a strong 
indication that the relative motion of the pair is linear (that is, just 
motion from an unrelated field star due to proper motion differences). The
{\bf H} indicates where the companion would have been at 1991.25, at the
Hipparcos epoch.}
\end{figure}


\section{Future Observing}

Due to the relatively even distribution of targets not yet observed by speckle 
interferometry, one observing run in each semester and each hemisphere will be necessary 
in order to observe all remaining exoplanet host stars. However, the target list for each 
of the four runs will be slight, less than one hundred stars each. With an approximately 
equal number of quality control and equatorial scale calibration pairs, each observing 
run could easily be completed in 2-3 nights. Priority would obviously be given to targets 
not observed before. Those observed by other speckle interferometric groups would be next
priority so they all have a common reduction algorithm. Figure 2 is an Aitoff plot of 
targets from the list of known exoplanet host stars taken from the
NStED\footnote{http://nsted.ipac.caltech.edu/cgi-bin/bgServices/nph-bgExec,
extracted 12 April 2011} database and gives their observation status.
 
\begin{figure}[t]
\begin{center}
{\epsfxsize 4.0in \epsffile{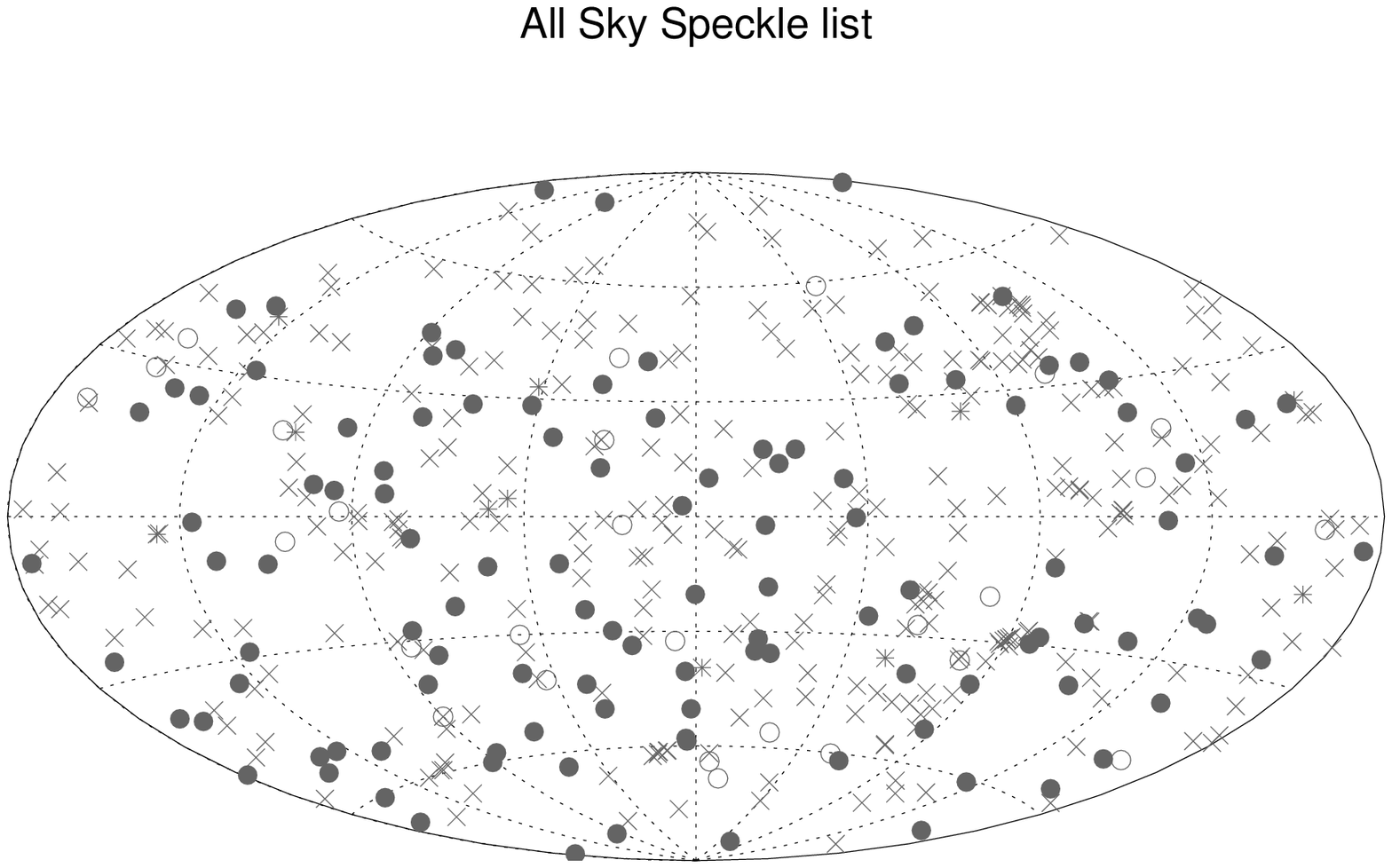}}
\end{center}
\caption{\small Aitoff projection of all 444 targets. Filled circles 
($N = 114$) are those listed in Table 2. Open circles ($N = 27$) are those
observed by CHARA or USNO with an ICCD and reduced with the DVA method. 
Asterisks ($N = 11$) are those observed by other interferometry groups, and
an $``$X" ($N = 292$) are those which have yet to be observed. A valid 
speckle measure is only counted if it was obtained on a 4m class telescope.}
\end{figure}

\acknowledgments

The USNO speckle interferometry program has been supported by NASA and
the SIM preparatory science program through NRA 98-OSS-007, SIM Key Project
MASSIF as well as No. NNH06AF701 issued through the Terrestrial Planet 
Finder Foundation Science program. Thanks are also provided to the U.S. 
Naval Observatory for their continued support of the Double Star Program. 
A portion of the research in this paper was carried out at the Jet 
Propulsion Laboratory, California Institute of Technology, under a contract 
with the National Aeronautics and Space Administration (NASA). This research
has made use of the SIMBAD database, operated at CDS, Strasbourg, France. 
Grateful acknowledgment is given to USNO interns Dean Kang, Laura Flagg and 
Ankit Patel for their processing of the speckle pixel data. The telescope 
operators and observing support personnel of KPNO and CTIO continue to 
provide exceptional support for visiting astronomers. Thanks to Alberto 
Alvarez, Skip Andree, Bill Binkert, Gale Brehmer, Bill Gillespie, Angel 
Guerra, Jim Hutchinson, Hillary Mathis, Oscar Saa, Patricio Ugarte, and the 
rest of the KPNO and CTIO staff. We also like to thank Richard Green of 
KPNO, who was able to provide us with two extra nights on the 4-m Mayall 
telescope during our January 2001 run. While we were hampered by poor 
weather, this additional allocation of time certainly helped us achieve a 
greatly enhanced completion fraction. A special thank you is also given to 
Hal Halbedel, who operated the telescope on all or part of each of these new
KPNO runs and was instrumental in the slit-mask work done at KPNO.


\begin{deluxetable}{ccl@{~}rcccccc}
\tabletypesize{\scriptsize}
\tablenum{1}
\tablewidth{0pt}
\tablecaption{New Interferometric Double}
\tablehead{
\colhead{Coordinates} & 
\colhead{WDS or} & 
\multicolumn{2}{c}{Discoverer} &
\colhead{HIP} & 
\colhead{BY} & 
\colhead{$\theta$} & 
\colhead{$\rho$} & 
\colhead{$\Delta$m} & 
\colhead{sep} \\
\colhead{$\alpha$,$\delta$ (2000)} & 
\colhead{$\alpha$,$\delta$ (2000)} & 
\multicolumn{2}{c}{Designation} & 
\colhead{~} & 
\colhead{2000.0+} & 
\colhead{($^{\circ}$)} & 
\colhead{($''$)} & 
\colhead{(mag)} &
\colhead{(AU)} \\
}
\startdata
01\phn26\phn08.78\phn$+$34\phn34\phn46.9 & 01262$+$3435 & WSI & 96 & 6702 & 1.0193 &    257.3 & 0.087 &     & 3.3 \\
                                         &              &     &    &      & 7.6049 & \phn78.3 & 0.109 & 2.3 & 4.2 \\
\enddata
\end{deluxetable}


\begin{deluxetable}{rrcrc}
\tablenum{2}
\tablewidth{0pt}
\tablecaption{Stars with No Companion Detected}
\tablehead{
\colhead{HIP \#} &
\colhead{HD \#} &
\colhead{Common} &
\colhead{BY} & 
\colhead{Telescope} \\ 
\colhead{~} & 
\colhead{~} & 
\colhead{Name} &
\colhead{2000.0+} & 
\colhead{~} \\
}
\startdata
   522 &    142\phm{B} & \nodata        &  1.5621 & C \\
  1292 &   1237\phm{B} & GJ 3021        &  1.5674 & C \\
  1499 &   1461\phm{B} & \nodata        &  1.5701 & C \\
  3391 &   4113\phm{B} & \nodata        &  1.5647 & C \\
  3479 &   4208\phm{B} & \nodata        &  1.5647 & C \\
  3497 &   4308\phm{B} & \nodata        &  1.5648 & C \\
  5054 &   6434\phm{B} & \nodata        &  1.5622 & C \\
  5529 &   7199\phm{B} & \nodata        &  1.5675 & C \\
  5806 &   7449\phm{B} & \nodata        &  1.5675 & C \\
  6379 &   7924\phm{B} & \nodata        &  1.0218 & K \\
  7513 &   9826\phm{B} & $\upsilon$ And &  1.0220 & K \\
  7599 &  10180\phm{B} & \nodata        &  1.5702 & C \\
  7978 &  10647\phm{B} & \nodata        &  1.5702 & C \\
  8159 &  10697\phm{B} & \nodata        &  1.0300 & K \\
  9683 &  12661\phm{B} & \nodata        &  1.0165 & K \\
 10138 &  13445\phm{B} & GJ 86          &  1.5676 & C \\
 10626 &  13931\phm{B} & \nodata        &  1.0219 & K \\
 12186 &  16417\phm{B} & \nodata        &  1.5676 & C \\
 12189 &  16246\phm{B} & 30 Ari         &  1.0304 & K \\
 12653 &  17051\phm{B} & $\iota$ Hor    &  1.5676 & C \\
 14954 &  19994\phm{B} & \nodata        &  1.0820 & C \\
 15323 &  20367\phm{B} & \nodata        &  1.0193 & K \\
 15527 &  20782\phm{B} & \nodata        &  1.0820 & C \\
 16537 &  22049\phm{B} & $\epsilon$ Eri & 10.0652 & C \\
 17096 &  23079\phm{B} & \nodata        &  1.0738 & C \\
 20723 &  28185\phm{B} & \nodata        &  1.0764 & C \\
 24681 &  34445\phm{B} & \nodata        &  1.0306 & K \\
 25110 &  33564\phm{B} & \nodata        &  1.0222 & K \\
 26381 &  37124\phm{B} & \nodata        &  1.0307 & K \\
 26394 &  39091\phm{B} & \nodata        &  1.0740 & C \\
 26394 &  39091\phm{B} & \nodata        &  6.1937 & C \\
 26664 &  37605\phm{B} & \nodata        &  1.0306 & K \\
 27887 &  40307\phm{B} & \nodata        &  1.0740 & C \\
 27887 &  40307\phm{B} & \nodata        &  6.1937 & C \\
 28767 &  40979\phm{B} & \nodata        &  1.0224 & K \\
 29804 &  43848\phm{B} & \nodata        &  1.0793 & C \\
 30034 &  44627\phm{B} & AB Pic         &  1.0740 & C \\
 30579 &  45364\phm{B} & \nodata        &  1.0794 & C \\
 30860 &  45350\phm{B} & \nodata        &  1.0198 & K \\
 30905 &  45652\phm{B} & \nodata        &  1.0308 & K \\
 31246 &  46375\phm{B} & \nodata        &  1.0767 & C \\
 31540 &  47186\phm{B} & \nodata        &  1.0794 & C \\
 32916 &  49674\phm{B} & \nodata        &  1.0224 & K \\
 32970 &  50499\phm{B} & \nodata        &  1.0794 & C \\
 33212 &  50554\phm{B} & \nodata        &  1.0279 & K \\
 33719 &  52265\phm{B} & \nodata        &  1.0823 & C \\
 36795 &  60532\phm{B} & \nodata        &  1.0795 & C \\
 37826 &  62509\phm{B} & \nodata        &  1.0199 & K \\
 38041 &  63765\phm{B} & \nodata        &  1.0742 & C \\
 38558 &  65216\phm{B} & \nodata        &  1.0742 & C \\
 40693 &  69830\phm{B} & \nodata        &  6.1911 & C \\
 40952 &  70642\phm{B} & \nodata        &  1.0743 & C \\
 43587 &  75732\phm{B} & 55 Cnc         &  1.0200 & K \\
 46076 &  81040\phm{B} & \nodata        &  1.0770 & C \\
 47007 &  82943\phm{B} & \nodata        &  1.0797 & C \\
 47202 &  83443\phm{B} & \nodata        &  1.0744 & C \\
 48235 &  85390\phm{B} & \nodata        &  1.0744 & C \\
 48739 &  86226\phm{B} & \nodata        &  1.0797 & C \\
 49699 &  87883\phm{B} & \nodata        &  1.0202 & K \\
 50473 &  89307\phm{B} & \nodata        &  1.0310 & K \\
 50921 &  90156\phm{B} & \nodata        &  6.1912 & C \\
 52521 &  93083\phm{B} & \nodata        &  1.0799 & C \\
 53721 &  95128\phm{B} & 47 UMa         &  1.0202 & K \\
 54906 &  97658\phm{B} & \nodata        &  1.0203 & K \\
 57172 & 101930\phm{B} & \nodata        &  1.0746 & C \\
 57291 & 102117\phm{B} & \nodata        &  1.0746 & C \\
 57370 & 102195\phm{B} & \nodata        &  1.0311 & K \\
 57443 & 102365\phm{B} & \nodata        &  6.1915 & C \\
 57443 & 102365\phm{B} & \nodata        & 10.0659 & C \\
 57931 & 103197\phm{B} & \nodata        &  1.0746 & C \\
 58451 & 104067\phm{B} & \nodata        &  6.1915 & C \\
 59610 & 106252\phm{B} & \nodata        &  1.0312 & K \\
 64295 & 114386\phm{B} & \nodata        &  1.0827 & C \\
 64426 & 114762\phm{B} & \nodata        &  1.0232 & K \\
 64457 & 114783\phm{B} & \nodata        &  5.1915 & C \\
 64459 & 114729\phm{B} & \nodata        &  1.0775 & C \\
 64459 & 114729\phm{B} & \nodata        &  1.0802 & C \\
 64924 & 115617\phm{B} & 61 Vir         &  6.1890 & C \\
 64924 & 115617\phm{B} & 61 Vir         &  8.4500 & K \\
 64924 & 115617\phm{B} & 61 Vir         & 10.0688 & C \\
 65721 & 117176\phm{B} & 70 Vir         &  1.0232 & K \\
 65721 & 117176\phm{B} & 70 Vir         &  6.1916 & C \\
 67275 & 120136\phm{B} & \nodata        &  1.0314 & K \\
 67275 & 120136\phm{B} & $\tau$ Boo     &  6.1916 & C \\
 71395 & 128311\phm{B} & \nodata        &  1.5664 & C \\
 71395 & 128311\phm{B} & \nodata        &  6.1916 & C \\
 72339 & 130322\phm{B} & \nodata        &  1.0829 & C \\
 74500 & 134987\phm{B} & \nodata        &  1.0830 & C \\
 74500 & 134987\phm{B} & \nodata        &  1.5611 & C \\
 77740 & 141937\phm{B} & \nodata        &  1.5666 & C \\
 78459 & 143761\phm{B} & $\rho$ CrB     &  8.4503 & K \\
 79242 & 142022A       & \nodata        &  1.5667 & C \\
 79248 & 145675\phm{B} & 14 Her         &  1.4986 & K \\
 80250 & 147018\phm{B} & \nodata        &  1.5667 & C \\
 80337 & 147513\phm{B} & \nodata        &  6.1919 & C \\
 83389 & 154345\phm{B} & \nodata        &  1.4960 & K \\
 83949 & 155358\phm{B} & \nodata        &  1.4961 & K \\
 86796 & 160691\phm{B} & $\mu$ Ara      &  1.5667 & C \\
 87330 & 162020\phm{B} & \nodata        &  1.5642 & C \\
 88348 & 164922\phm{B} & \nodata        &  8.4506 & K \\
 90004 & 168746\phm{B} & \nodata        &  1.5614 & C \\
 90485 & 169830\phm{B} & \nodata        &  1.5614 & C \\
 91085 & 171238\phm{B} & \nodata        &  1.5614 & C \\
 94075 & 178911B       & \nodata        &  1.4990 & K \\
 94645 & 179949\phm{B} & \nodata        &  1.5614 & C \\
 96901 & 186427\phm{B} & 16 Cyg         &  1.4991 & K \\
 97336 & 187123\phm{B} & \nodata        &  1.4990 & K \\
 97546 & 187085\phm{B} & \nodata        &  1.5670 & C \\
 98505 & 189733\phm{B} & \nodata        &  8.4614 & K \\
 98767 & 190360\phm{B} & \nodata        &  8.4563 & K \\
 99711 & 192263\phm{B} & \nodata        &  5.8680 & K \\
 99825 & 192310\phm{B} & GJ 785         &  1.5616 & C \\
101806 & 196050\phm{B} & \nodata        &  1.5615 & C \\
101966 & 196885\phm{B} & \nodata        &  8.4481 & K \\
104903 & 202206\phm{B} & \nodata        &  1.5618 & C \\
106006 & 204313\phm{B} & \nodata        &  1.5618 & C \\
106353 & 204941\phm{B} & \nodata        &  1.5618 & C \\
108375 & 208487\phm{B} & \nodata        &  1.5672 & C \\
108859 & 209458\phm{B} & \nodata        &  8.4615 & K \\
109378 & 210277\phm{B} & \nodata        &  1.5645 & C \\
111143 & 213240\phm{B} & \nodata        &  1.5618 & C \\
112441 & 215497\phm{B} & \nodata        &  1.5646 & C \\
113137 & 216437\phm{B} & \nodata        &  1.5647 & C \\
113238 & 216770\phm{B} & \nodata        &  1.5646 & C \\
113357 & 217014\phm{B} & 51 Peg         &  7.5883 & K \\
113357 & 217014\phm{B} & 51 Peg         &  8.4617 & K \\
116727 & 222404\phm{B} & $\gamma$ Cep   &  1.0218 & K \\
116727 & 222404\phm{B} & $\gamma$ Cep   &  1.4993 & K \\
116906 & 222582\phm{B} & \nodata        &  1.5674 & C \\
\enddata
\end{deluxetable}


\vfill\eject

\appendix
\section {Additional Measures of Known Pairs}

Due to the high incidence of single stars among the exoplanet hosts, a 
substantial number of double stars were observed contemporaneously with the 
exoplanet host observations to ensure that the observing conditions and 
detection capabilities given above were met. Additional measures of known or
suspected doubles were made as time permitted. Table A1 lists 549 mean 
positions for 485 known systems. Column (1) is the WDS identification 
(arcminute coordinate), Column (2) lists the Discovery Designation, and 
Column (3) provides the Hipparcos number. Column (4) gives the epoch of 
observation, and Columns (5) and (6) provide the relative astrometry.  
Column (7) contains the notes for these systems. Also found in in table are ten pairs 
resolved for the first time and sixty pairs which are here confirmed; estimated magnitude
differences for the new pairs (when available) are listed in the Notes column.

\pagestyle{empty}

\scriptsize
\begin{deluxetable}{ll@{~}r@{~}lrrrrc}
\tablenum{A1}
\tablewidth{0pt}
\tablecaption{Other Pairs Measured}
\tablehead{
\colhead{WDS or} & 
\multicolumn{3}{c}{Discovery} &
\colhead{HIP} & 
\colhead{BY} & 
\colhead{$\theta$} & 
\colhead{$\rho$} & 
\colhead{Note} \\
\colhead{$\alpha$,$\delta$ (2000)} & 
\multicolumn{3}{c}{Designation} & 
\colhead{~} & 
\colhead{2000.0+} & 
\colhead{($^{\circ}$)} & 
\colhead{($''$)} & 
\colhead{~} \\
}
\startdata
00024$+$1047 & A   & 1249 & AB    &    190 & 7.5994 &  74.8\phm{:} & 0.135\phm{:} &     \\
00026$-$0829 & A   &  428 &       &    210 & 7.6019 & 344.5\phm{:} & 0.174\phm{:} &     \\
00055$-$1835 & RST & 3340 &       &    --- & 1.5701 & 287.4\phm{:} & 0.303\phm{:} &     \\
00095$+$1907 & COU &  247 &       &    768 & 7.5992 & 259.1\phm{:} & 0.294\phm{:} &     \\
00117$+$6145 & TDS & 1338 &       &    --- & 7.6074 & 111.2\phm{:} & 0.527\phm{:} & 1   \\
00167$+$3629 & STT &    4 &       &   1336 & 7.6021 &  99.7\phm{:} & 0.232\phm{:} &     \\
00174$+$0853 & A   & 1803 & AB    &   1392 & 7.6019 & 307.2\phm{:} & 0.145\phm{:} &     \\
00182$+$7257 & A   &  803 &       &   1461 & 7.6021 & 307.9\phm{:} & 0.262\phm{:} &     \\
\enddata
\tablenotetext{1}{~Confirming observation.}
\tablenotetext{2}{~Two measurements are averaged to give this mean position.}
\tablenotetext{3}{~Calibration system.}
\tablenotetext{4}{~Previously known as RST3558a.}
\tablenotetext{5}{~Confirmed with HRCam on SOAR 4.2m, CTIO 4m (Tokovinin et al.\ 2010).}
\tablenotetext{6}{~Three measurements are averaged to give this mean position.}
\tablenotetext{7}{~First measure of newly resolved pair. Primary is HD 341480. }
\tablenotetext{8}{~Confirming observation or more likely a new component.}

\end{deluxetable}


\begin{references}

\reference {} Bagnuolo, W.G., Jr., Mason, B.D., Barry, D.J., Hartkopf, W.I.,
              \& McAlister, H.A. 1992, AJ 103, 1399

\reference {} Duquennoy, A. \& Mayor, M. 1991, A\&A 248, 485

\reference {} ESA 1997, The Hipparcos and Tycho Catalogues, ESA SP-1200

\reference {} Farrington, C.D. et al.\ 2012 ({\it in preparation})

\reference {} Harrington, R.S. 1981, Planetary orbits in multiple star 
              systems, in Life in the Uuniverse, Cambridge, MIT Press, 119

\reference {} Hartkopf, W.I. et al.\ 2000, AJ 119, 3084

\reference {} Hartmann, M., Guenther, E.W. \& Hatzes, A.P. 2010, ApJ 717,
              348

\reference {} Jao, W.-C., Mason, B.D., Hartkopf, W.I., Henry, T.J. \& Ramos,
              S.N. 2009, AJ 137, 3800

\reference {} Mason, B.D., Hartkopf, W.I., Gies, D.R., Henry, T.J., \&
              Helsel, J.W. 2009, AJ 137, 3358

\reference {} Mason, B.D., Hartkopf, W.I. \& Wycoff, G.L. 2011, AJ ({\it in
              press})

\reference {} Mason, B.D., Wycoff, G.L., Hartkopf, W.I., Douglass, G.G. \&
              Worley, C.E. 2001, AJ 122, 3466\footnote{{\it The Washington 
              Double Star Catalog}, published in {\it Second USNO Double
              Star CD 2006.5} and available online at {\tt 
              http://ad.usno.navy.mil/wds/wds.html}.}

\reference {} Raghavan, D., Henry, T.J., Mason, B.D., Subasavage, J.P., Jao,
              W.-C., Beaulieu, T.D. \& Hambly, N.C. 2006, ApJ 646, 523

\reference {} Raghavan, D. et al.\ 2010, ApJS 190, 1

\reference {} Roberts, L.C., Jr., Turner, N.H., ten Brummelaar, T.A., Mason,
              B.D. \& Hartkopf, W.I. 2011, ({\it in progress})

\reference {} Tokovinin, A., Mason, B.D. \& Hartkopf, W.I. 2010, AJ 139, 743

\end{references}
\end{document}